\documentclass[12pt]{article}

\usepackage{scicite}

\usepackage{times}
\usepackage{caption}
\usepackage{graphicx}
\usepackage{color}

\newenvironment{sciabstract}{%
\begin{quote} \bf}
{\end{quote}}

\newcounter{lastnote}

\title{Quantum-well states at the surface of the heavy fermion superconductor URu$_2$Si$_2$}
\author
{Edwin Herrera,$^{1,2,3}$ Isabel Guillam\'on,$^{3}$ V\'ictor Barrena,$^{3}$ \\ William Herrera,$^{2}$ Jose Augusto Galvis,$^{1}$ Alfredo Levy Yeyati,$^{4}$ \\ Jan Rusz,$^{5}$ Peter M. Oppeneer,$^{5}$ Georg Knebel,$^{6}$ \\ Jean Pascal Brison,$^{6}$ Jacques Flouquet,$^{6}$ Dai Aoki,$^{7}$ Hermann Suderow$^{3}$\\
\\
\normalsize{$^{1}$Facultad de Ingenier\'ia y Ciencias B\'asicas}\\
\normalsize{Universidad Central, 111321 Bogot\'a, Colombia}\\
\normalsize{$^{2}$Departamento de F\'isica}\\
\normalsize{Universidad Nacional de Colombia, 111321 Bogot\'a, Colombia}\\
\normalsize{$^{3}$Laboratorio de Bajas Temperaturas, Unidad Asociada UAM/CSIC}\\
\normalsize{Departamento de F\'isica de la Materia Condensada}\\
\normalsize{Instituto Nicol\'as Cabrera and Condensed Matter Physics Center (IFIMAC)}\\
\normalsize{Universidad Aut\'onoma de Madrid, E-28049 Madrid, Spain}\\
\normalsize{$^{4}$Departamento de F\'isica Te\'orica de la Materia Condensada}\\
\normalsize{Instituto Nicol\'as Cabrera and Condensed Matter Physics Center (IFIMAC)}\\
\normalsize{Universidad Aut\'onoma de Madrid, E-28049 Madrid, Spain}\\
\normalsize{$^{5}$Department of Physics and Astronomy, Uppsala University}\\
\normalsize{Box 516, S-75210 Uppsala, Sweden}\\
\normalsize{$^{6}$University Grenoble Alpes, CEA, IRIG-PHELIQS, F-38000 Grenoble, France}\\
\normalsize{$^{7}$IMR, Tohoku University, Oarai, Ibaraki 311 - 1313, Japan}\\
}

\date{}

\begin{document} 


\baselineskip24pt


\maketitle 

\clearpage

\begin{sciabstract}
Electrons can form a two-dimensional electron gas at metal surfaces, where lateral confinement leads to quantum-well states. Such states have been observed for highly itinerant electrons, but it remains an open question whether quantum-well states can be formed from strongly correlated electrons. Here we study atomically flat terraces on surfaces of the heavy-fermion superconductor URu$_2$Si$_2$ using millikelvin scanning tunneling spectroscopy. We observe two-dimensional heavy fermions (2DHF) with an effective mass 17 times the free electron mass that form quantized states separated by a fraction of a meV. Superconductivity at the surface is induced by the bulk into the 2DHF. Our results provide a new route to realize quantum well states in correlated quantum materials.
\end{sciabstract}

Two-dimensional electronic states confined at surfaces are ubiquitously observed within the gaps of the bulk dispersion relation of metals opened by the periodic crystalline lattice \cite{Tamm1932,Shockley1939,Hasegawa1993,Crommie1993,ECHENIQUE1989,Li1998}. These states can be further confined through scattering by features forming closed geometries at the surface, for example at a terrace between two surface steps. Then, quantized energy levels emerge, typically from itinerant electron states of the two-dimensional surface band\cite{Crommie1993,Wenderoth1999,Burgi1999}. Their separation in energy is often of hundreds of meV, i.e. a few percent of the width of the surface band. Their lifetime (or the level width) is set by the interaction between the surface and bulk states and is slightly smaller than the level separation\cite{Pines1966,Giuliani2005,Burgi1999,Kirchmann2010,Seo2010}.  Quantum well states form at the surface of many different materials and provide a fascinating ground to visualize electron correlations and study the influence of geometry or of adatoms and impurities\cite{Knorr2002,Li1998,Ternes09,Fiete2003,Manoharan2000,Fiete2001}. However, it remains difficult to obtain quantum-well states whose separation in energy is in the meV range and thus comparable to bulk low energy phenomena, such as for example superconductivity. Increasing the spatial extension of the confinement potential could eventually lead to a reduction of the level separation. But the lifetime would then remain of the same order and the level width would increase well above the level separation. To reduce the level separation to the meV range one needs systems with a narrow band crossing the Fermi level. So called heavy-fermion metals provide such a flat band in the bulk due to collective Kondo hybridization. As a consequence of the increased density of states, these systems are highly susceptible to present exotic low temperature properties, as for example d-wave or triplet p-wave superconductivity\cite{hewson_1993,Lohneysen2007,ColemanBook07,Flouquet2005,Patil2016,Boariu2010,Zhang2018,Pirie2020,Fujimori2021,Denlinger2022}. It was found that reducing the dimensionality in heavy fermions leads to suppression of magnetism and to a deviation from standard Fermi liquid electronic interactions in certain superlattices of compounds with a layered crystalline structure\cite{doi:10.1126/science.1183376}. Furthermore, narrow surface bands with a Dirac dispersion were found in a semiconducting heavy fermion\cite{Pirie2020}. However, two-dimensional heavy fermions (2DHF) have not been observed at surfaces of metallic compounds, and no quantum-well states due to lateral confinement have been realized. Here we show that 2DHF appear at the surface of a heavy fermion metal and that these form quantum well states at terraces between steps.

Bulk URu$_2$Si$_2$ exhibits correlated heavy electron bands crossing the Fermi level in the bulk (Fig.\,1(A)) and shows a transition to a mysterious phase (called hidden order, HO) characterized among other features by the opening of a partial gap in the bandstructure below $T_{HO}=$17.5 K, out of which an unconventional superconducting state develops below $T_{c}=$1.5 K \cite{Oppeneer11,Mydosh2011,Mydosh2014}. Its surface has been intensively investigated; there are electronic surface states with bands whose effective masses are mostly relatively small \cite{Boariu2010,Zhang2018,Denlinger2022,Fujimori2021}. Here we use Scanning Tunneling Microscopy (STM) to investigate small-sized atomically flat terraces on the surface of URu$_2$Si$_2$. We observe 2DHF with an effective mass 17 times the free electron mass and quantization due to lateral confinement.

We present a STM image of terraces in URu$_2$Si$_2$ in Fig.\,1(B). The bias voltage dependence of the tunneling conductance coincides with results obtained in previous STM work on flat surfaces, except for the features associated to the 2DHF discussed below \cite{Schmidt2010,Yuan2012,Aynajian2010,Hamidian11,Herrera2021}. The shape and temperature dependence of the tunneling conductance shows band hybridization induced by electronic correlations and the formation of the heavy fermion state when cooling (Fig.\,1(A)). In Fig.\,1(C,D,E) we show results at three different temperatures (details provided in Supplementary Sections S1 and S2). At temperatures above $T_{HO}$ (Fig.\,1(C)) we observe a strong asymmetric Fano lineshape due to interfering simultaneous tunneling into localized f-electron and conduction electron states. Below $T_{HO}$ (Fig.\,1(D)), a gap opens around the Fermi level. Inside the gap, we observe a van Hove anomaly at $\varepsilon_{vH}\approx -1.5$ meV due to the dispersion bending of the low energy bandstructure. These features were also found in previous work\cite{Schmidt2010,Yuan2012,Aynajian2010,Hamidian11,Herrera2021,Morr2016}. When cooling further (Fig.\,1(E)), below the superconducting critical temperature $T_c=1.5$ K, we find a peak at $\varepsilon_{+}\approx 0.4$ meV and the superconducting gap with the value expected from $T_c$ and BCS theory, $\Delta_{SC} \approx 200$ $\mu$eV. We analyze in more detail the latter two features below.

To discuss the observation of the quantized 2DHF, let us focus on the tunneling conductance along the white line shown in Fig.\,1(B) for very low bias voltages, between $\pm$1.6 mV. Along this line, we identify four terraces of different sizes, $L_1\approx2$ nm, $L_2\approx5.5$ nm, $L_3\approx20$ nm and $L_4\approx38.5$ nm (Fig.\,1(F)). We observe peaks in the tunneling conductance at $\varepsilon_{vH}\approx -1.5$ meV and at $\varepsilon_{+}\approx 0.4$ meV all over the flat regions of the terraces (orange and blue arrows in Fig.\,1(G)). At the steps, these features move into an edge state which we discuss below. There is a strong bias voltage and position dependence of the tunneling conductance which is different for each terrace. We show in Fig.\,1(H) representative tunneling conductance curves at each terrace. For the larger terraces,  $L_3$ and $L_4$, we can identify a set of new peaks (vertical arrows in Fig.\,1(H)).

To further understand these peaks, let us focus on the terrace $L_3$, marked by a dashed rectangle in Fig.\,1(B). We symmetrize the tunneling spectroscopy along the long axis of the terrace and show the resulting tunneling conductance in Fig.\,2(A). We can identify a set of peaks in the tunneling conductance which evolve both in position and bias voltage. Removing the background introduced by the features at $\varepsilon_{vH}=-1.5$ meV and at $\varepsilon_{+}=0.4$ meV from the tunneling conductance (Fig.\,1(E)) leads to the pattern shown in Fig.\,2(B). We calculate the expected position of peaks from lateral quantization using interfering wavefunctions partially reflected at steps. The quantization pattern for confined electrons ressembles the Fabry-P\'erot expression for an interferometer made by partially reflecting mirrors\cite{Burgi1998}, with the reflection coefficient $r$ and the phase shift $\phi$ as free parameters (details in the Supplementary Section S3). The results are the white dots in Fig.\,2(A,B), whose position coincides well with the peaks in the conductance pattern observed in the experiment. In Fig.\,2(C) we plot as circles the position of the peaks as a function of the wavevector $k$ and as a line the dispersion relation $E=E_0 + \hbar^2 k^2/(2m^*)$, where $m^*$ is the effective mass and $E_0$ the bottom of the band. We obtain $E_0 =-2.3$ meV and $m^*=17 m_0$, with $m_0$ the free electron mass, i.e., a massive electron state. A detailed comparison of the tunneling conductance vs. position with the square of wavefunctions confined by a lateral potential leads to excellent fits, shown in Fig.\,2(D). The phase shift $\phi$ determines the position and energy of the peaks (white dots in Fig.\,2(B)) and the best account of our observations is obtained with $\phi=-\pi$. We find values around $r\approx$ 0.2, which slightly increase when approaching $E_0$ (Fig.\,2(E)). The low value of $r$ is also found in simple metals, for example $r$ is between 0.2 and 0.4 in Ag and Cu \cite{Burgi1998,Pendry1994,PhysRevLett.73.1015}. However, the energy dependence (Fig.\,2(E)) is here completely different than in usual metals. While $r$ varies mildly in Cu or Ag in the range of a few eV (see discontinous line in Fig.\,2(E) and Refs.\,\cite{Burgi1998,Pendry1994,PhysRevLett.73.1015}), here we observe instead that $r$ decreases markedly in a range of a few meV (points in Fig.\,2(E)). We can reproduce the observed dependence of $r$ vs. bias voltage assuming a rectangular well of $L=$20 nm size and a barrier with a height of 5 meV (blue line in Fig.\,2(E))\cite{Burgi1998,Pendry1994}. It is also insightful to trace the tunneling conductance as a function of the bias voltage at the center of the terrace (Fig.\,2(F)) and compare it to the expectation for $r \approx 1$ (dashed line in Fig.\,2(F)) and for $r=0.4$ (continuous line in Fig.\,2(F)). We see that both the periodicity and shape are well explained by a reduced $r$ in an energy range of a few meV, i.e.\ two orders of magnitude below the energy range observed in normal metals\cite{Burgi1998,Pendry1994,PhysRevLett.73.1015,Hasegawa1993,Crommie1993,ECHENIQUE1989,Li1998,Gartland1975,Kevan1992}.

To further investigate the quantized levels, we have fitted each peak (see black arrows in Fig.\,2(B)) to a Lorentzian function, whose width $\Gamma$ (Fig.\,2(B)) provides the lifetime $\tau$ (Fig.\,2(G)) of the quantum well states. Taking a 2D electron gas, we expect $1/\tau \propto [(E-E_0)/E_0]^2[\ln |(E-E_0)/E_0|-\ln(2q_{TF}/k_F)-1/2]$ with $q_{TF}=0.0906 $\AA$^{-1}$ the Thomas-Fermi screening length and $k_F$ the Fermi wavevector \cite{Giuliani1982} (dashed line in Fig.\,2(G)). Our data are not well reproduced by this expression. Instead taking $1/\tau = [\Gamma_0 + (E_0/\pi)[(E-E_0)/E_0]^2]^{-1}$ with $E_0=-2.3$ meV and $\Gamma_0 \approx 60 \, \mu$eV (line in Fig.\,2(G)) we find a much better account of our experiment\cite{Seo2010,Giuliani1982}. This shows that the lifetime $\tau$ is set by the decay into bulk states. Quantum well states sense the bulk Fermi liquid correlations, given by the quadratic energy term in $1/\tau$. This has been observed in surface states of noble metals, monolayers of Pb, and in Sb. However, in those cases the energy range was three orders of magnitude above the one we discuss here\cite{Pines1966,Giuliani2005,Burgi1999,Kirchmann2010,Seo2010}.

Using $\Gamma_0\approx 0.06$ meV, obtained from the fit, we estimate the lifetime of the ground state as $\tau_0(E_0=-2.3$ meV$)=\hbar/\Gamma(0)\approx 11$ ps. Similarly, the lifetime of states close to the Fermi level is $\tau(E=0)=\hbar/\Gamma(E=0)\approx 3$ ps. We can also estimate a value for a mean free path, $\ell_0=v_F \hbar/\Gamma(0) \approx 0.14$ $\mu m$, taking for $v_F$ the Fermi velocity of URu$_2$Si$_2$. This value is of the same order of magnitude as those observed in ultraclean URu$_2$Si$_2$ single crystals reported in Ref.\,\cite{Kasahara2009}. Furthermore, the finite $\Gamma_0$ is related to the reflection coefficient $r$, also due to electronic interactions with the bulk. The blue line in Fig.\,2(E) assumes a single isolated quantum well and thus perfect reflection $r=1$ at $E_0$. In the presence of many terraces, as we find in our experiment, we can consider multiple scattering events from different terraces\cite{Seo2010,Heller1994}. Then, in the extrapolation of $r$ to $E_0$, $\Gamma_0$ is given by $\Gamma_0=(\hbar / 2m^*)[(1-r_0)/2Lr_0]^2$, where $L$ is the terrace width\cite{Seo2010}. We find $r_0\approx 0.25 \pm 0.05$, which is consistent with the extrapolation of $r(E)$ to $E=E_0=-2.3$ meV from our data (dotted line in Fig.\,2(E)).

To vindicate the heavy-fermion surface state we have performed Density Functional Theory (DFT) calculations of the surface bandstructure of a slab of URu$_2$Si$_2$ (Supplementary Section S4). We find a shallow, U derived f-electron band with a flat dispersion relation compatible to our experiments around the $X$ point of the simple tetragonal Brillouin zone. The bulk electronic spectrum is gapped in this part of the Brillouin zone\cite{Denlinger2022,Fujimori2021}. The rest of the Brillouin zone provides surface states with much smaller effective masses.

It is important to see how the 2DHF is modified at steps and impurities. We discuss the behavior at steps here (providing more details provided in Supplementary Section S5). The behavior at impurities is provided in Supplementary Section S6. In the conductance map shown in Fig.\,1(G) we find that the features in the tunneling conductance change their position in bias voltage as a whole when approaching a step. The peak due to the van Hove anomaly at $\varepsilon_{vH}=-1.5$ meV and the peak at $\varepsilon_{+}=0.4$ meV both vanish close to the steps. Instead we find a high peak at $\varepsilon_{1DES}=-0.35$ meV (see for example the red blobs at the sides of Fig.\,2(A)). This indicates that the lateral quantized levels evolve at the step into 1D-Edge States (1DES) of the 2DHF. An 1DES is the result of charge accumulation at the step and a modified local electron density\cite{Smoluchowski1941,Avouris1994,Namba1993}. A careful characterization of the $\varepsilon_{1DES}$ feature at terraces of different sizes shows that the 1DES is always at the same distance from the step and has the same lateral size (Supplementary Section S5). Furthermore, a gap opens at the step and is filled at  $\varepsilon_{1DES}$ by the resonant 1DES.

This precise and succesful comparison between the observed tunneling conductance and all relevant features of a 2DHF, including quantized levels sensing the bulk correlations and the formation of edge states at steps, provides a neat picture of the surface of the heavy fermion URu$_2$Si$_2$---there are 2DHF at the surface which are connected to the bulk.

The question is now how does this situation impacts on the features in the tunneling conductance due to superconductivity. When we look on the energy range below the superconducting gap, $\Delta=200$ $\mu$eV, we remark that the superconducting features are strongly suppressed, with a large zero bias conductance. There are indications for unconventional superconductivity in bulk URu$_2$Si$_2$, for instance with d-wave symmetry order parameter \cite{Kasahara2009}. This can contribute to the suppression of the superconducting features, but it hardly leads to the small superconducting features observed in our experiment. Similar small sized superconducting features are found in other heavy fermion superconductors, such as CeCoIn$_5$, CeCu$_2$Si$_2$ or UTe$_2$, and remain difficult to explain\cite{Allan13,PhysRevB.93.045123,Jiao2020}. Most notably, macroscopic measurements as specific heat or thermal conductivity provide in all these systems a negligible zero temperature extrapolation, suggesting that the density of states at the Fermi level is very small \cite{hewson_1993,Lohneysen2007,ColemanBook07,Flouquet2005,Aoki2019}. As we have seen, the 2DHF is strongly coupled to the bulk and is thus superconducting by proximity. The coupling should take into account the features in the density of states induced by quantum well states and by the features at $\varepsilon_{vH}$ and $\varepsilon_{+}$. The latter two have a particularly strong influence in the measured tunneling conductance curves. To explain our results, we have considered a one dimensional transport scheme between tip and sample, where tunneling occurs simultaneously into the 2DHF ($t_1$, Fig.\,3(A)) and the features observed in the bulk (we take $\varepsilon_{vH}$, $t_2$, Fig.\,3(A)). We couple the 2DHF to $\varepsilon _{vH}$ ($t_{vH}$, Fig.\,3(A)), to $\varepsilon_{+}$ ($t_+$, Fig.\,3(A)) and to the superconducting bulk ($t_s$, Fig.\,3(A), details given in the Supplementary Section S7). We find that the zero bias tunneling conductance is very large, mostly due to the large width of the features at $\varepsilon _{vH}$ and $\varepsilon _{+}$. With this model, we fit well the major features of the tunneling conductance and can follow the temperature dependence of the tunneling conductance (Fig.\,3(C,D). This solves the discrepancy between macroscopic and surface experiments found in many heavy fermion superconductors, and shows the relevance of multiple couplings between low energy resonances, surface states and bulk superconductivity to understand the tunneling spectroscopy of strongly correlated superconductors.

In summary, we have observed 2DHF in terraced surfaces inside the HO phase of URu$_2$Si$_2$. The 2DHF exhibits quantum well states when confined between steps with energy separation of fractions of meV. This discovery opens new possibilities to study the interplay of quantized 2DHF states and unconventional superconductivity, since a number of heavy fermion materials display unconventional superconductivity in the bulk, often within other strongly correlated phases. Apart from URu$_2$Si$_2$, there is e.g.\  CeCu$_2$Si$_2$, CeCoIn$_5$, UBe$_{13}$, UPt$_3$ and UTe$_2$. The proposed superconducting states for these systems are singlet d-wave or triplet p- and f-wave states. These could exhibit 2DHF and the associated edge states could incorporate excitations with unique properties such as Majorana fermions following non-abelian statistics. Furthermore, since the 2DHF states are quantized by lateral confinement, manipulation of correlated quantum confined states is thus in principle possible by atomic engineering, or by simply looking at terraces with different geometries. This opens new avenues to generate, isolate, and manipulate excitations in unconventional superconductors.

\paragraph*{Acknowledgments.} This work was supported by the Spanish State Research Agency (PID2020-114071RB-I00, CEX2018-000805-M), by the Comunidad de Madrid through program nanomag-cost-cm (Program No.\ S2018/NMT-4321), and by EU (PNICTEYES ERC-StG-679080 and COST NANOCOHYBRI CA16218). H.S., E.H.\ and I.G.\ acknowledge SEGAINVEX at UAM for design and construction of STM cryogenic equipment. J.A.G.\ thanks for the support of the Ministerio de Ciencia, Tecnología e Innovación de Colombia (Grant  No.\ 122585271058). J.R.\ and P.M.O.\ acknowledge support from the Swedish Research Council (VR) and from the Swedish National Infrastructure for Computing (SNIC), through Grant No\ 2018-05973.

\paragraph*{Author contributions.} E.H.\ performed the study and made all experiments, with the supervision of I.G.. E.H.\ analyzed the data and compared with theory with the supervision of W.H.\ and J.A.G. J.R.\ and P.M.O.\ performed the band structure calculations and discussed the features of surface states. The models to account for superconducting features and for the behavior at the step were proposed by W.H.\ and A.L.Y.. D.A.\ synthesized, prepared and characterized the samples with support from G.K.\ and J.P.B.\ for their characterization. D.A., J.F.\ and H.S.\ proposed the study. The manuscript was written by E.H., I.G., W.H., A.L.Y and H.S.\ with contributions from all authors.

\bibliographystyle{Science}

\clearpage

\begin{figure}
	\begin{center}
	\centering
	\includegraphics[width=0.98\textwidth]{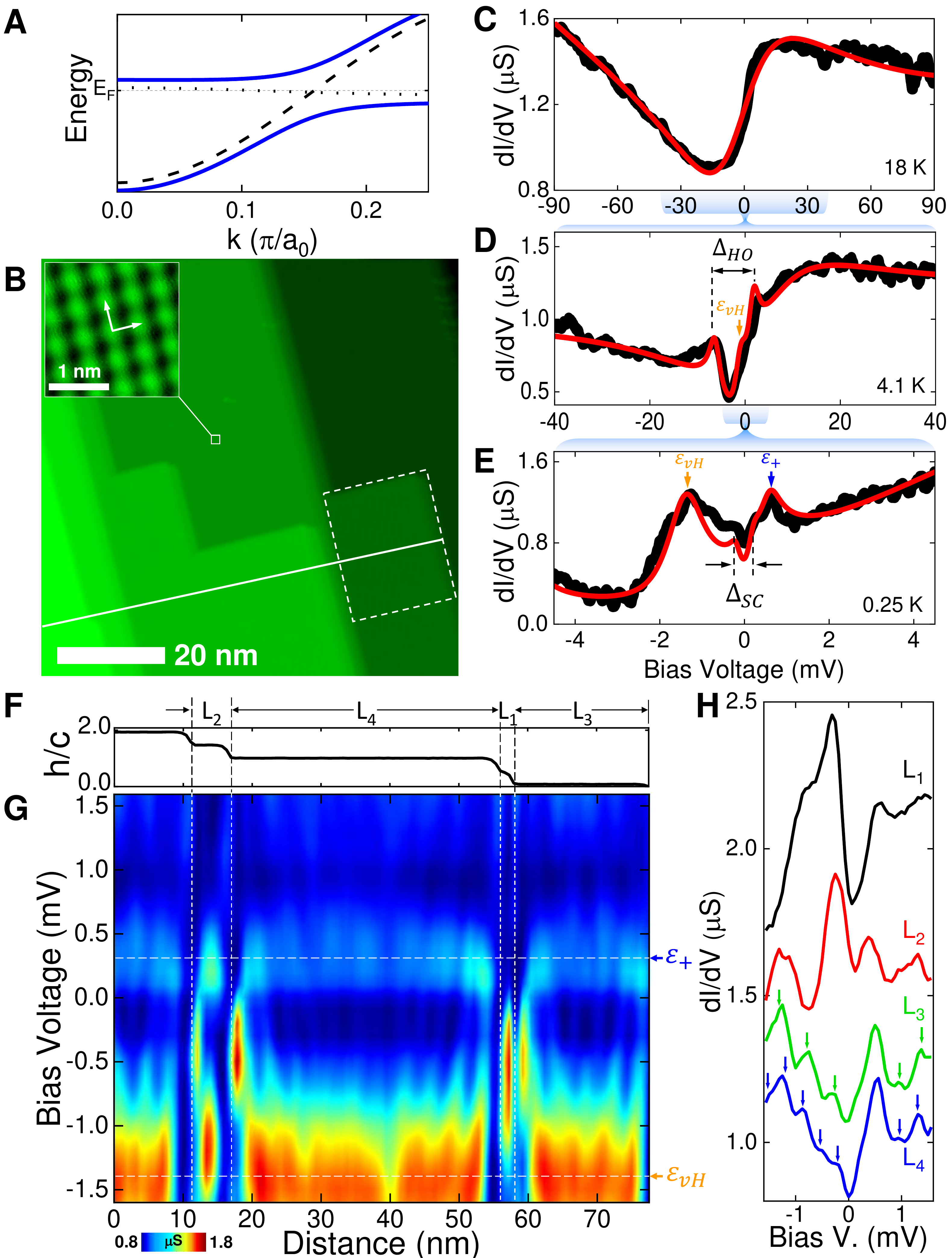}
	\end{center}
	\vskip -0.5 cm
	\caption{}
	\label{Fig1}
\end{figure}

\clearpage
\noindent
{\bf Fig.\,1 Tunneling conductance on URu$_2$Si$_2$ terraces.} ({\bf A}) Simplified representation of the low temperature hybridization process leading to heavy fermions. The dashed (dotted) lines show the unhybridized light (heavy) electron bands at high temperatures. Full lines show the resulting low temperature hybridized bands. ({\bf B}) STM topography image of several terraces on URu$_2$Si$_2$ separated by $c/2$. Inset shows a zoom into the region at the white square, and unveils the U atomic surface lattice (arrows indicate the in-plane crystalline directions). Dashed rectangle represents the field of view on which we focus in in Fig. 2. ({\bf C-E}) In black are the tunneling conductance curves at  temperatures $T>T_{HO}$, $T<T_{HO}$ and $T<T_c$ respectively. Red lines are the fits to the model described in the Supplementary Sections S3 and S7. We highlight the size of the gap opening in the HO phase ($\Delta_{HO}$). ({\bf F}) Height profile along the white line in ({\bf B}) identifying four different terraces of sizes $L_1$ to $L_4$. ({\bf G}) Tunneling conductance as a function of the bias voltage along the while line in ({\bf B}). ({\bf H}) Representative tunneling conductance curves obtained on each terrace. Vertical arrows mark the positions of peaks in $L_3$ and $L_4$. Orange and blue arrows at $\varepsilon_{vH}$ and $\varepsilon_+$ in ({\bf D}, {\bf E}, {\bf G}) identify the van Hove feature inside the HO gap and an additional feature resulting from hybridization in the bandstructure. 

\clearpage

\begin{figure}
	\begin{center}
	\centering
	\includegraphics[width=1\textwidth]{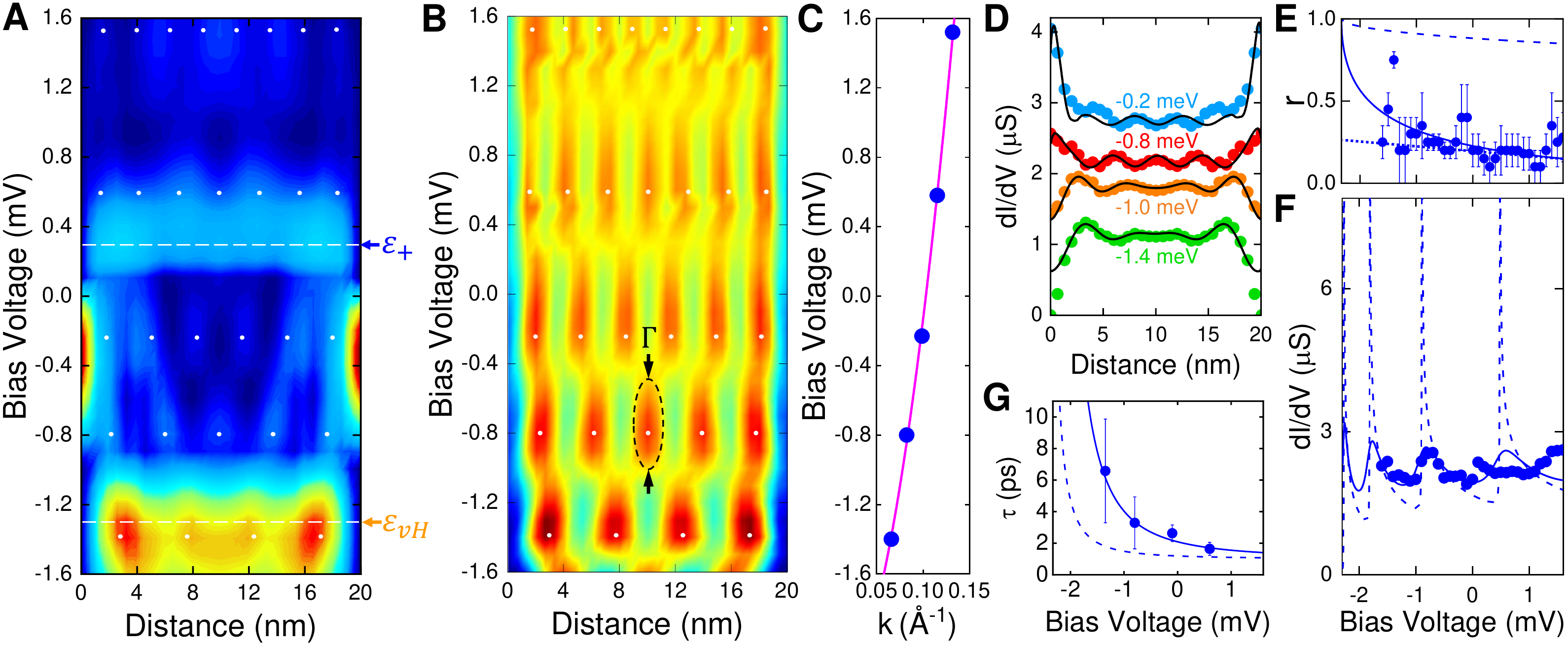}
	\end{center}
	\vskip -0.5 cm
	\caption{{\bf 2DHF and electron-in-a-box quantization.} ({\bf A}) The tunneling conductance at the $L_3$ terrace is shown by a color scale as a function of the distance. Dashed white lines and horizontal arrows show $\varepsilon_{vH}$ (orange) and $\varepsilon_{+} $ (blue). ({\bf B}) The same data as in ({\bf A}), with a subtracted background, described in the Supplementary Section S3. White dots in ({\bf A}) and ({\bf B}) mark the position of peaks in the conductance of the electron-in-a-box model discussed in the text. Black dotted ellipse shows the width of the quantized levels, $\Gamma$, described in the text. ({\bf C}) Circles show the reciprocal space position of the white dots in ({\bf A}) and ({\bf B}). Magenta line provides the electron dispersion relation with $m^*=17$ $m_0$, being $m_0$ the free electron mass. ({\bf D}) Lines show the results of the electron-in-a-box calculations and circles the tunneling conductance. ({\bf E}) The dashed blue line is $r$ obtained in Cu(111), from Ref.\,\cite{Pendry1994}. The continuous blue line is the calculated $r$ assuming an effective mass $m^*=17$ $m_0$. Circles are the results obtained from the experiment. The dotted line takes into account the finite width $\Gamma$ of the quantization peaks. ({\bf F}) Tunneling conductance as a function of the bias voltage is shown by blue circles at the center of the terrace. Dashed blue lines are for perfect reflection, $r=1$, and continuous lines for $r\approx$ 0.4. ({\bf G}) Circles show the lifetime of the quantum well states, $\tau$ as a function of the bias voltage. The dashed blue line is the expectation for a two-dimensional electron gas and the continuous line describes quantum states whose width is set by their interaction with the heavy quasiparticles of the bulk.}
	\label{Fig2}
	\end{figure}

\clearpage

\begin{figure}
	\begin{center}
	\centering
	\includegraphics[width=0.8\textwidth]{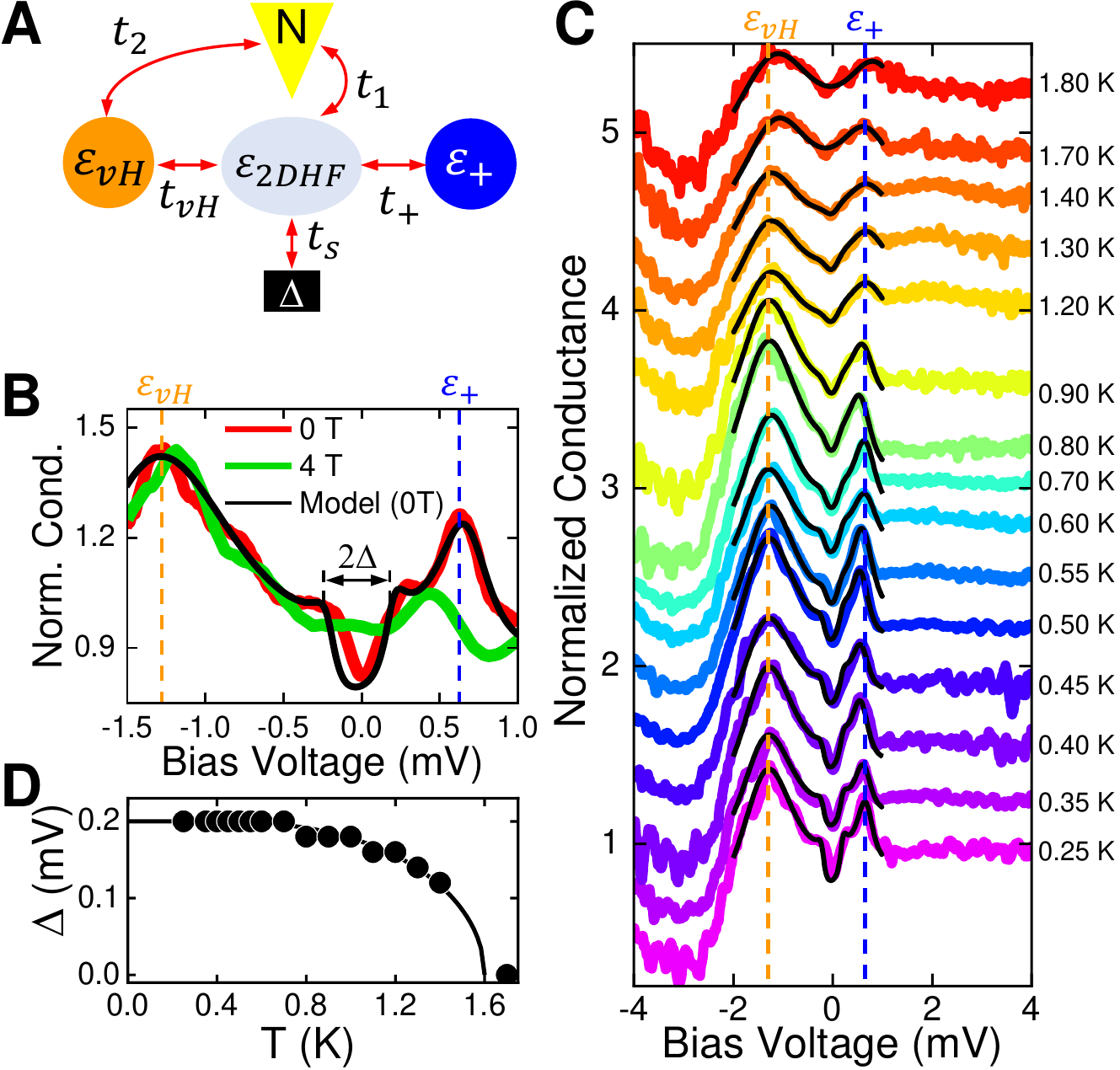}
	\end{center}
	\vskip -0.5 cm
	\caption{{\bf Superconductivity in the 2DHF.} ({\bf A}) We schematically model the observed tunneling spectroscopy using a normal tip (yellow) which is coupled to the bulk and the 2DHF. We consider the 2DHF as a broadened level (light blue ellipse) that couples to $\varepsilon_{vH}$ (orange circle) and $\varepsilon_{+}$ (blue circle). The 2DHF couples to the superconductor by proximity. ({\bf B}) Tunneling conductance vs bias voltage at zero field (red) and at 4 T (green). The features at $\varepsilon_{vH}$ and $\varepsilon_+$ are marked by, respectively, orange and blue dashed lines. The fit at zero magnetic field is shown by a black line. The superconducting gap $\Delta$ is marked by an arrow. ({\bf C}) Tunneling conductance vs temperature at zero field. Curves are shifted for clarity. Black lines are fits to the model schematically shown in ({\bf A}) and described in Supplementary Section\,S7. ({\bf D}) Temperature dependence of the superconducting gap $\Delta(T)$ is shown by black circles. The BCS temperature dependence (black line) is shown as a guide.}
	\label{Fig3}
	\end{figure}

\end{document}



\baselineskip24pt


\maketitle 

\clearpage
\renewcommand{\contentsname}{This file includes:}
\begin{spacing}{0.01}
  \tableofcontents
  \setcounter{tocdepth}{0}
\pagenumbering{gobble}
\end{spacing}

\section*{Section S1. Materials, Methods and measured surface termination}
\label{sec1}
\addcontentsline{toc}{section}{\normalfont Section S1. Materials, Methods and measured surface termination}

Single crystals of URu$_2$Si$_2$ were grown by the Czochralski technique in a tetra-arc furnace. We have scanned samples for a low residual resistivity and a high critical temperature, close to 1.5 K. Such samples were then cut in bar shape with dimensions $ 4 \times 1 \times 1 $ mm$^3$ with the long distance parallel to the c-axis. We mounted the samples on the sample holder of a Scanning Tunneling Microscope (STM). The STM was mounted in a dilution refrigerator (base temperature 100 mK). The setup has an resolution in energy below 10 $\mu$eV and leads to clean BCS s-wave tunneling conductance curves in Al and other s-wave superconductors\cite{Suderow2011,doi:10.1063/5.0059394}. Further details, including image rendering software, is provided in Refs.\,\cite{doi:10.1063/5.0064511,doi:10.1063/1.2432410}. The STM head includes a low temperature movable sample holder that allows to cleave the sample at cryogenic temperatures\cite{Suderow2011,Herrera2021}.

\begin{figure}
	\begin{center}
	\centering
	\includegraphics[width=1\textwidth]{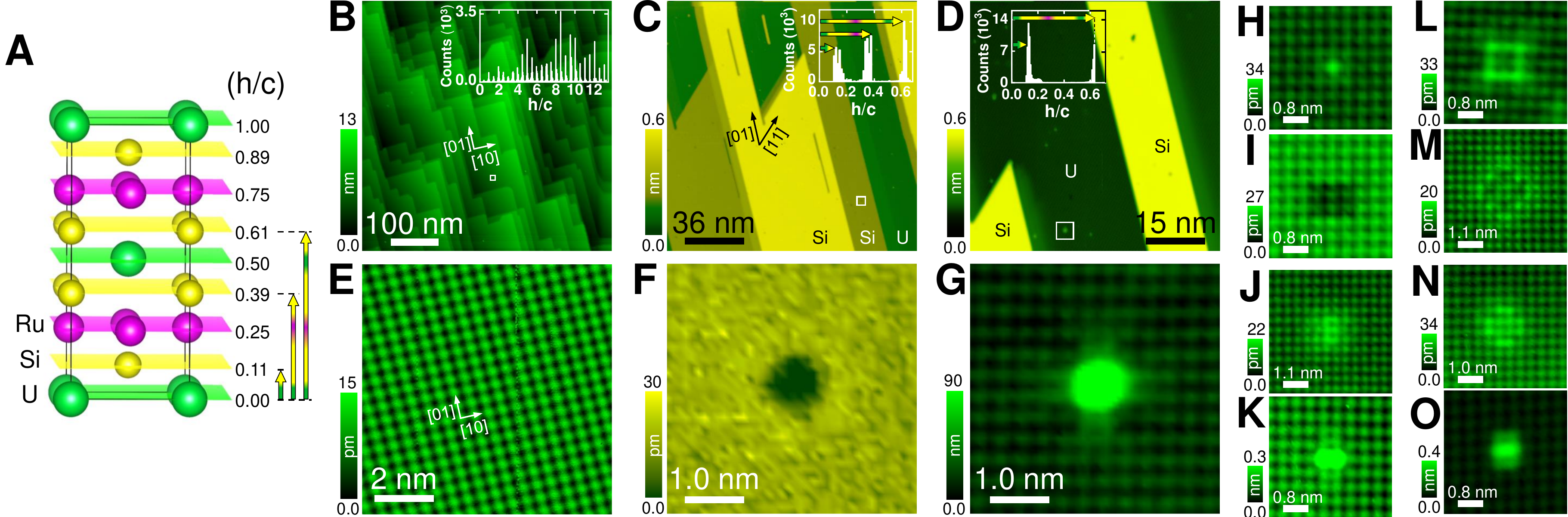}
	\end{center}
	\vskip -0.5 cm
	\caption{{\bf Different surface terminations in URu$_2$Si$_2$.} ({\bf A}) URu$_2$Si$_2$ crystal structure. U atoms are shown in green, Ru in magenta and Si in yellow. We show using the same colors the corresponding planes and indicate the distances between planes, normalized to the c-axis lattice parameter. With colored arrows we highlight the distance between the planes observed in the STM images. ({\bf B}) STM topogaphy image at a surface with square shaped terraces. As we discuss in the text, these are all U-terminated terraces. Height scale is given by the bar on the left. In the upper right inset we show a height histogram. Notice that all peaks are located at an integer of $c/2$. The crystal axes are shown as white arrows. The small white square is the area plotted in (E), see below. ({\bf C-D}) STM topography images at surfaces showing terraces with a triangular shape. These are distinct terraces. The crystal axes are shown as arrows. In the upper right inset we show the height histogram, with distances between terraces having different sizes. Colored arrows are as in ({\bf A}) and identify height differences between U and Si terraces. The small white square provides the area scanned in ({\bf F}), see below. In ({\bf D}) we show another set of triangular shaped terraces and the corresponding height histogram in the upper left inset. The small white square provides the field of view shown in ({\bf G}). We show other kinds of defects in {\bf H} to {\bf O}, with the color scale given by the bars on the left of each image. All data were taken at 100 mK with a tunneling current $I_{tunnel}$=10 nA and a bias voltage $V_{Bias}$=10 mV.}
	\label{FigS1}
	\end{figure}

We focus on U-terminated surfaces. In Fig.\,S\ref{FigS1}(A) we show the URu$_2$Si$_2$ crystal unit cell highlighting the U-, Si- and Ru-planes; their inter-layer distances are indicated in units of the c-axis lattice parameter. In Figs.\,S\ref{FigS1}(B-D) we show STM images corresponding to different surface terminations. These surfaces are all obtained after cryogenic cleaving. On the surfaces full of square shaped terraces we find the results obtained in the text. An example is shown in Fig.\,S\ref{FigS1}(B). All the observed terraces (Fig.\,S\ref{FigS1}(A,B)) are separated by $c/2 \approx 4.84$ \AA. In Fig.\,S\ref{FigS1}(C,D) we show terraces with a triangular shape, where we do not observe the phenomena discussed in the main text. Here the distance between consecutive terraces is either $\sim0.11c$, $\sim0.39c$ or $\sim0.61c$, which correspond, respectively, to the three possible U-Si planes (arrows in Fig.\,S\ref{FigS1}(A)) . Therefore, we see that the surfaces with terraces having a triangular shape correspond to Si layers, with sometimes a U layer in between. By contrast, the surfaces with terraces having a square shape are U terminated. Atomically resolved images inside terraces having a square shape (Figs. S\ref{FigS1}(E)) provide the square atomic U lattice with an in-plane constant lattice of $a = 4.12$ \AA (Fig. S\ref{FigS1}(E)). In Fig. S\ref{FigS1}(F) we show a typical atomic size image on Si-terminated surfaces. We do not observe atomic resolution and have sometimes circular defects. Defects in the U-terminated surfaces are very different, as shown in Figs.\,S\ref{FigS1}(G-O). We distinguish two distinct types of defects. The defects can be either point like protusions (Figs.\,S\ref{FigS1}(G,H)) or troughs (Figs.\,S\ref{FigS1}(I)). Sometimes, defects are arranged in small size square or rectangular structures (Figs.\,S\ref{FigS1}(J,K,L,M,N,O)). Most of these defects are probably due to vacancies or interstitial atoms in layers below the U surface layer.

\section*{Section S2. Fano anomaly and density of states at temperatures above and below the HO transition.}
\label{sec2}
\addcontentsline{toc}{section}{\normalfont Section S2. Fano anomaly and density of states at temperatures above and below the HO transition.}

We now explain the red lines in Fig.\,1(C-E) of the main text. These correspond essentially to results obtained previously in Refs.\,\cite{Schmidt2010,Aynajian2010}. The tunneling conductance is due to simultaneous tunneling into heavy and light bands, as in other heavy fermion compounds. The red line in Fig.1(C) of the main text for $T=18 K$ follows a Fano function 

\begin{equation}
g(E)=A\frac{(q +(E-E_0)/\Gamma_F)^2}{(E-E_0)/\Gamma_F+1},
\label{EqFano}
\end{equation}

where $A$ is a constant of proportionality, $q$ is the ratio between two tunneling paths, $E_0$ is the Fano resonance energy with width $\Gamma_F=2 \sqrt{(\pi k_BT)^2+2(k_BT_K)^2}$, $T_K$ being the Kondo temperature \cite{Schmidt2010,Aynajian2010}.
For the fit we include an asymmetric linear background due to the degree of particle‐hole asymmetry in the light conduction band \cite{Schmidt2010,Figgins2010}. To account for the thermal broadening we convolute the result with the derivative of the Fermi-Dirac distribution. We find $q =0.8 \pm 0.5$, $E_0=3 \pm 1 mV$, $\Gamma_F = 22 \pm 1 mV$ and $T_K = 90 \pm 5 K$; all consistent with previous reports \cite{Schmidt2010,Aynajian2010}.   

Inside the HO phase (red line in Fig.\,1(D) of the main text) we use the same Fano function, multiplied by an asymmetric BCS-like gap function with an offset $\delta_E$, $g_{ HO}=(E-\delta_E-i\gamma_{HO})/[\sqrt{(E-\delta_E-i\gamma_{HO})^2-\Delta_{HO}^2}]$. The resulting function is convoluted with the derivative of the Fermi-Dirac distribution function. We find $\delta_E(4.1 K) =1.5 \pm 0.5 meV$, $\Delta_{HO} (4.1 K)=4.0 \pm 0.5 meV$; all consistent with previous reports  \cite{Schmidt2010,Aynajian2010}. Notice that we also observe further features at lower temperatures and smaller bias voltages (red line in Fig.\,1(E) of the main text). For these, we include Lorentzian functions centered at $\varepsilon_{vH}$ and $\varepsilon_{+}$ and an asymmetric background. The Fano lineshape, the HO gap opening and the van Hove feature $\varepsilon_{vH}$ have been found and discussed previously\cite{Schmidt2010,Yuan2012,Aynajian2010,Hamidian11,Morr2016}. The small feature at $\varepsilon_{+}$ occurs at a very similar energy range as a kink in the band structure measured in Th-doped URu$_2$Si$_2$\cite{Hamidian11,Morr2016}. In addition to those previous results, we determine the 2DHF and their quantized states and discuss the consequences for the observation of superconductivity at the surface.

\section*{Section S3. Quantum well states at terraces in between steps.}
\label{sec3}
\addcontentsline{toc}{section}{\normalfont Section S3. Quantum well states at terraces in between steps.}

\begin{figure}
	\begin{center}
	\centering
	\includegraphics[width=0.7\textwidth]{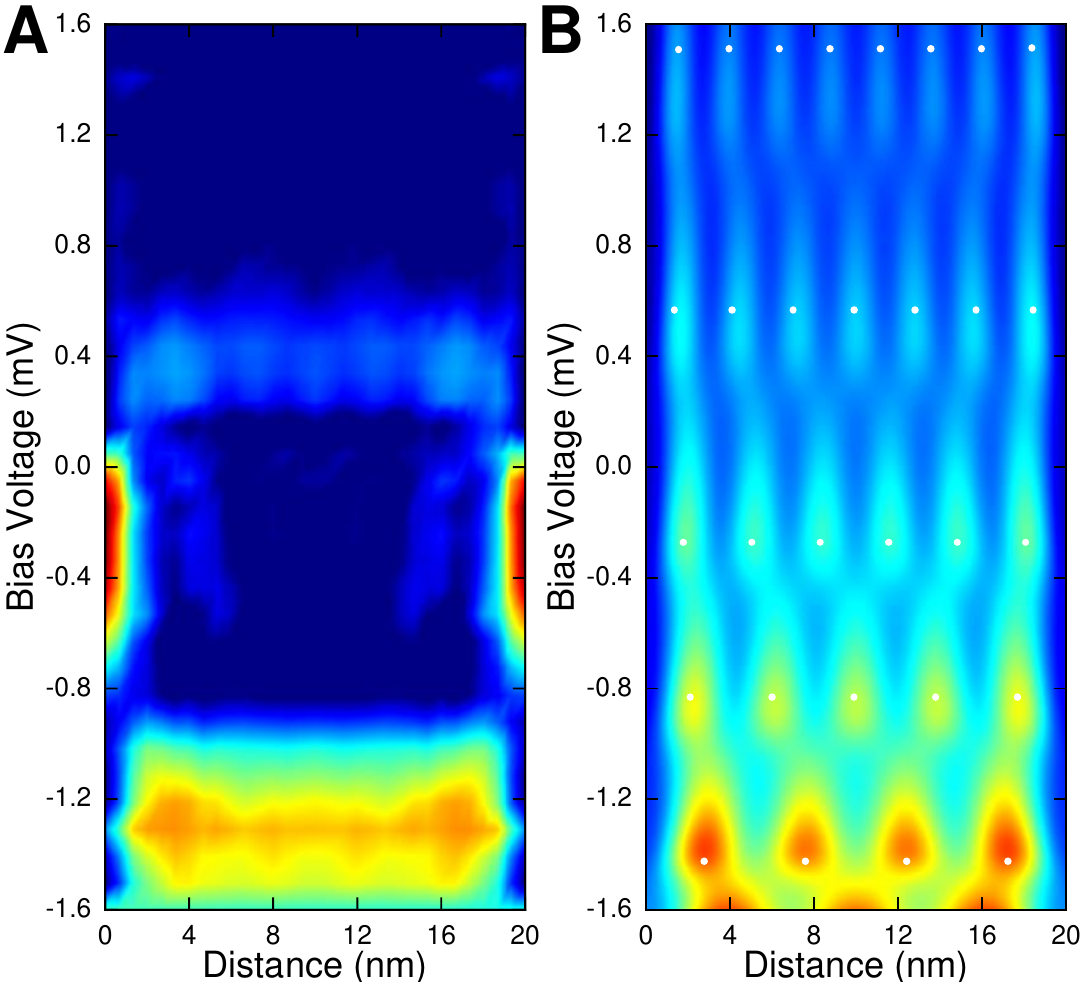}
	\end{center}
	\vskip -0.5 cm
	\caption{{\bf Subtracted tunneling conductance background and quantized density of states pattern.} ({\bf A}) Bias voltage dependence of the subtracted tunneling conductance background vs distance on a color scale from blue (low conductance) to red (high conductance). ({\bf B}) Tunneling conductance calculated using the parameters discussed in the text. Quantized levels are represented by white points.}
	\label{FigS2}
	\end{figure}

Fig.\,S\ref{FigS2}(A) shows the experimental tunneling conductance background subtracted to the Fig.\,2(A) to obtain the Fig.\,2(B). To model the quantum well states we use the Fabry-P\'erot interferometer expression for the density of states $g_{FP} (x,E)$ given by 

\begin{multline}
g_{FP} (x,E)  =  C \frac{L_0}{\pi}\int^k_0 \frac{dq}{\sqrt{k^2+q^2}}\frac{1}{1+r^4 -2r^2cos(2qL+2\phi)}\\
\times ((1-r^2)[1+r^2+2rcos(2q(x-L)-\phi)]\\
+(1-r^2)[1+r^2+2rcos(2qx+\phi)])
\label{Eq1}
\end{multline}

with $k=\sqrt{2m^*(E-E_0)/\hbar^2}$, $m^*$ the electronic effective mass, $r$ the reflection amplitude, $\phi$ the phase and $L$ the width of the terrace \cite{Burgi1998}. The Fabry-P\'erot interferometer is an optical resonator made of semi-reflecting mirrors and provides a simple and insightful way to model electronic wave functions confined between two wells. We assume a symmetric potential well with $L=20$ nm, $r=0.5$ and $\phi=-\pi$. The pattern generated by Eq.\,\ref{Eq1} is shown in Fig.\,S\ref{FigS2}(B). White points provide the positions of quantized levels and are at the same positions as the white points in Fig.\,2(A-B) of the main text. The black lines in Fig.\,2(D) of the main text are fits to the Eq.(\ref{Eq1}). To account for the behavior at the edges, we add the Eq.(\ref{Eq2}) for the one dimensional edge state (1DES) discussed below. We use the parameters extracted for the terrace $L_3$, discussed in Table S\ref{Tab1}.

\begin{figure}
	\begin{center}
	\centering
	\includegraphics[width=0.8\columnwidth]{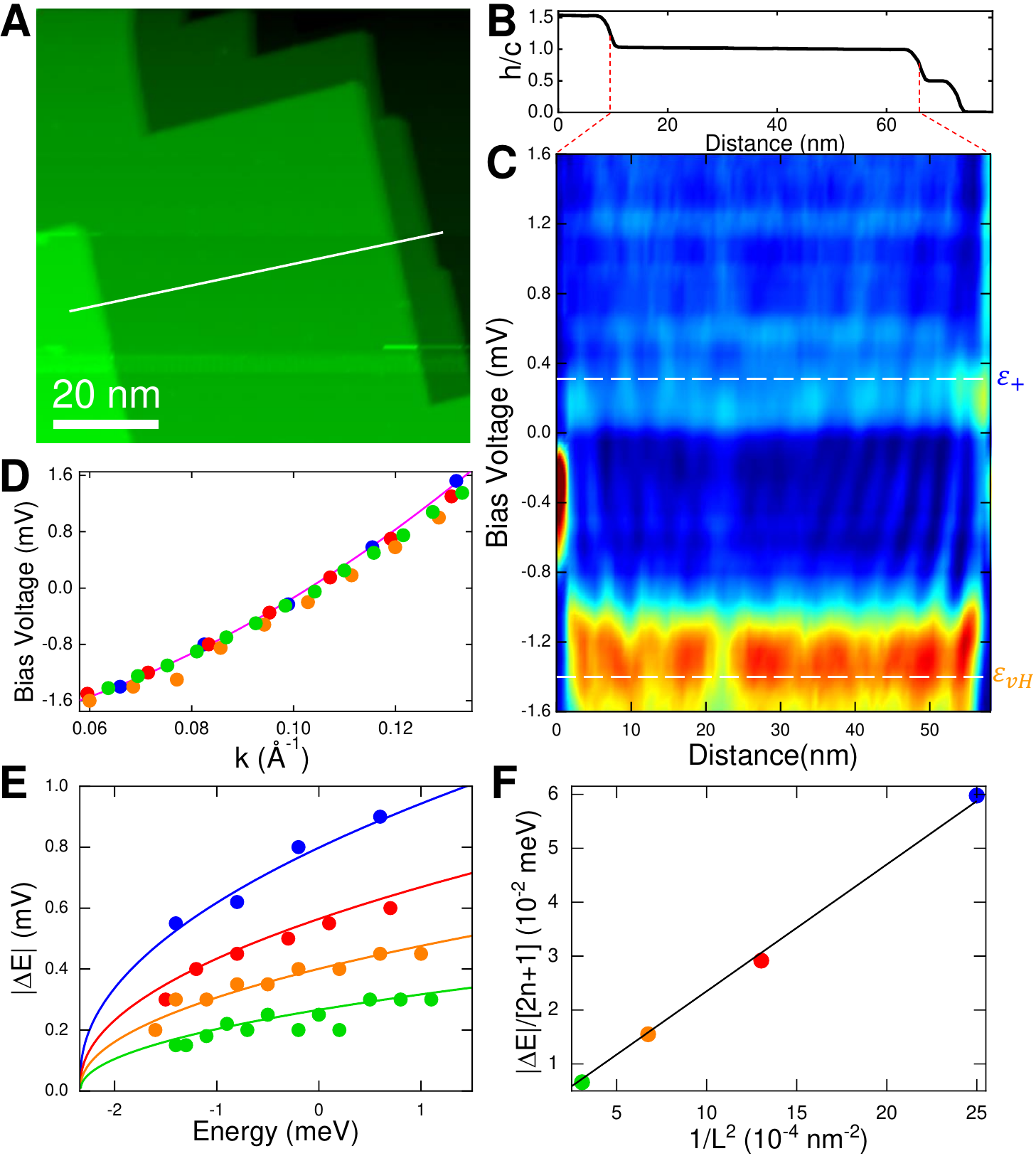}
	\end{center}
	\vskip -0.5 cm
	\caption{{\bf Dispersion relation and quantization on different terraces.} In ({\bf A}) we show a STM image on a field of view containing a few U-terminated terraces. In ({\bf B}) we show the height profile along the white dashed line in ({\bf A}). In ({\bf C}) we show the tunneling conductance along the portion indicated in ({\bf B}) as a color scale from red (high conductance) to blue (low conductance). We mark the position of the features at $\varepsilon_+$ and $\varepsilon_{vH}$ with white dashed lines. In ({\bf D}) we show the dispersion relation of the 2DHF as a magenta line. Circles in blue, red, orange and green are the positions of the quantized levels obtained from different terraces as described in the text (size $L$ of each terrace is of 20 nm blue, 28 nm red, 38.5 nm orange and 57 nm green). In ({\bf E}) we show as colored dots (color code as in ({\bf D})) the difference between energy levels $\Delta E$ as a function of the energy. Lines are a square root dependence, from $\Delta E = E_{n+1}-E_n=\left(\frac{\hbar^2\pi^2}{2m^*L^2}\right)\left( 2n+1 \right)$. In ({\bf F})  we show the average of $\Delta E$ (color code as in ({\bf D})) as a function of the length of the terrace $L$.
}
	\label{FigS3}
	\end{figure}

The 2DHF quantization was observed on the surfaces of different URu$_2$Si$_2$ samples. In Fig.\,S\ref{FigS3} we show the result on another sample. Notice that here terraces have different sizes. We show in Fig.\,S\ref{FigS3}(A) the STM topography image. In Fig.\,S\ref{FigS3}(B) we show a height profile through the white line in Fig.\,S\ref{FigS3}(A). In Fig.\,S\ref{FigS3}(C) we represent the tunneling conductance along the same profile. We observe similar tunneling conductance curves as those presented in Fig.\,1(G) of the main text. Notice the features at $\varepsilon_+$ and $\varepsilon_{vH}$. The quantized levels are also readily observed. These occur, however, at different energy values as the size of the terrace $L$ is different than in the main text. In Fig.\,S\ref{FigS3}(D) we represent the values of the quantized levels found in terraces of different sizes $L$ by different symbols; we show the dispersion relation of the 2DHF as a magenta line.

In Fig.\,S\ref{FigS3}(E) we show as colored dots the bias voltage dependence of the energy spacing $\Delta E$ between consecutive quantized levels for terraces $L_3$, $L_4$ (Fig.\ref{FigS1}(F,G)), the terrace shown in Fig.\,S\ref{FigS3}(C) and a terrace with length $L=27$ nm (not shown). We can write that $\Delta E = E_{n+1}-E_n=\left(\frac{\hbar^2\pi^2}{2m^*L^2}\right)\left((n+1)^2-n^2\right)$, with $n=1,2,3, ...$.  This gives a square root dependence of $\Delta E$ on the energy, shown in Fig.\,S\ref{FigS3}(E). In Fig.\,S\ref{FigS3}(F) we plot the average value of $\frac{\Delta E}{2n+1}$ for each terrace as a function of $L$. We find the expected $\frac{1}{L^2}$ dependence.

\clearpage

\begin{figure}
	\begin{center}
	\centering
	\includegraphics[width=0.7\textwidth]{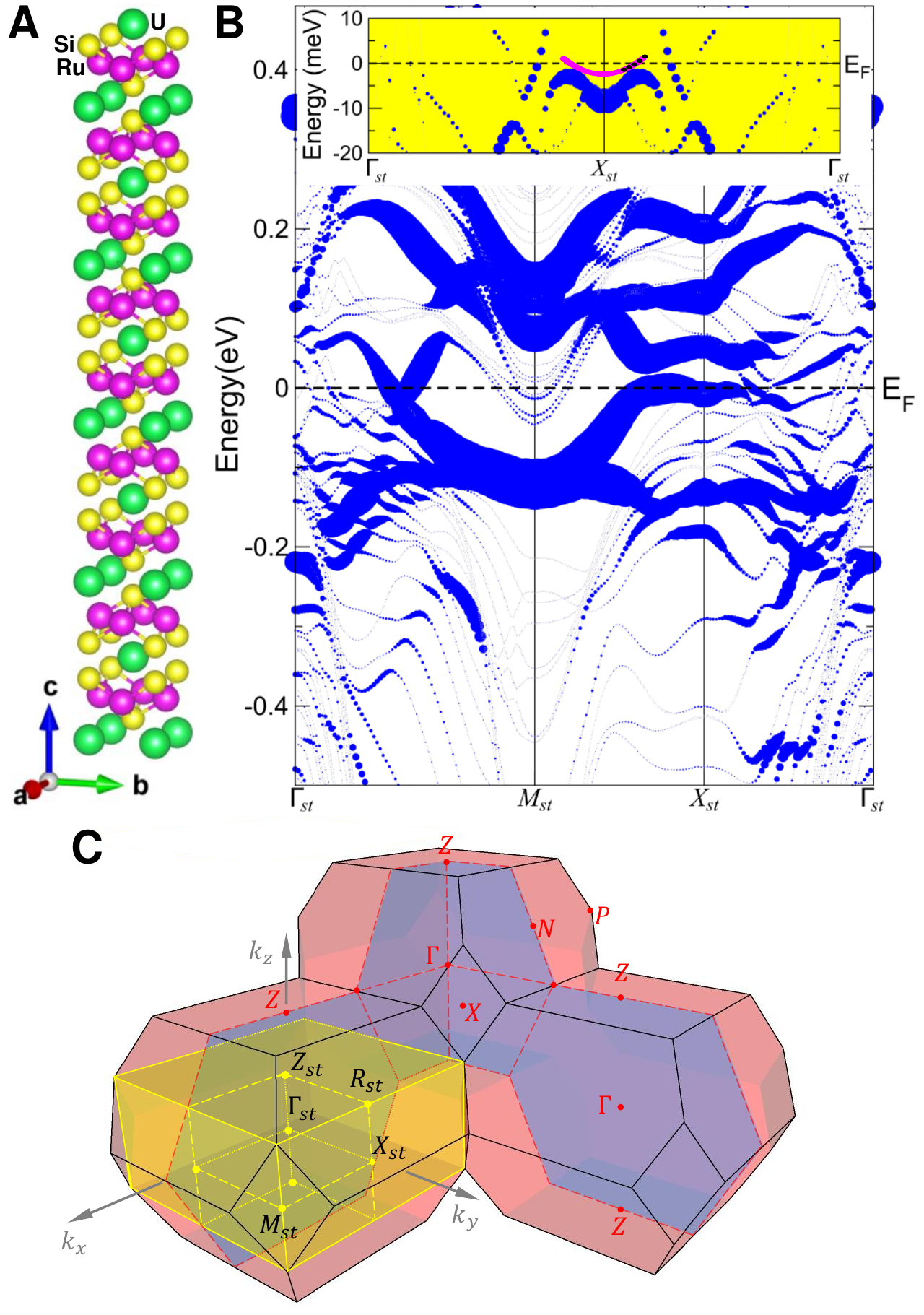}
	\end{center}
	\vskip -0.5 cm
	\caption{{\bf DFT calculations of the surface band structure around the X point of the simple tetragonal Brillouin zone.} ({\bf A}) U-terminated supercell structure of URu$_2$Si$_2$ used for DFT calculations. ({\bf B})  Bandstructure of URu$_2$Si$_2$ in a slab calculation (blue points), described in the text, along the high symmetry directions of the simple tetragonal Brillouin zone, $\Gamma_{st}$, $M_{st}$ and $X_{st}$. Size of points provides the U spin up character of the bands. In the upper inset we show a zoom around the $X_{st}$ point. The magenta line provides the dispersion relation compatible with our experiments, and the black points the quantized levels we identified (from $n=4$ to $n=8$). ({\bf C}) The usual Brillouin zone construction of URu$_2$Si$_2$, with the tetragonal Brillouin zone (red lines) and the simple tetragonal (st) construction (yellow lines) used to described the low temperature HO phase.}
	\label{FigS4}
	\end{figure}

\section*{Section S4. Band structure calculations at U-terminated surfaces of URu$_2$Si$_2$.}
\label{sec4}
\addcontentsline{toc}{section}{\normalfont Section S4. Band structure calculations at U-terminated surfaces of URu$_2$Si$_2$.}

The band structure of bulk URu$_2$Si$_2$ has been analyzed previously in detail using density functional calculations (DFT) \cite{PhysRevB.82.205103,Elgazzar09,Ikeda12}. Relevant results coincide with angular resolved photoemission, STM and quantum oscillation studies\cite{Aoki2012,Ohkuni1999,Yoshida2010,Kawasaki2011,Meng13,Bareille2014,SantanderSyro2009,Denlinger2022}.

Several surface states have been observed by ARPES \cite{Fujimori2021,Zhang2018,Boariu2010,Denlinger2022}. The surface state discussed in Ref.\,\cite{Boariu2010,Zhang2018} is formed by a hole-like band with its maximum at -35 meV, and is thus far from what we observe here. At the X point of the Brillouin zone, there are no bulk states. ARPES measurements show hints of surface-like bands with 2D character at these points \cite{Fujimori2021}. We have taken a closer look on the $X$ point through DFT calculations. To this end, we have built a U-terminated supercell consisting of thirty seven atomic layers, giving a total of ten U layers (Fig.\,S\ref{FigS4}(A)). We have performed density functional theory calculations using the full-potential linearized augmented plane waves method with local orbitals as implemented in the WIEN2k package \cite{Wien2020}. Atomic spheres radii were set to 2.5, 2.5 and 1.9 Bohr radii for U, Ru and Si, respectively. We have used $19 \times 19 \times 1$ mesh of $\mathbf{k}$-points in the first Brillouin zone, reduced by symmetry to 55 distinct $\mathbf{k}$-points. $RK_{max}$ parameter was set to 6.5, resulting in a basis size of approximately 5400 (over 100 basis functions per atom). Spin-orbital coupling has been included in the second variational step \cite{Kunes2001} and relativistic local orbitals were included for U $6p_{1/2}$ and Ru $4p_{1/2}$ states. The basis for calculations of spin-orbital eigenvalue problem consisted of scalar-relativistic valence states of energies up to 5$\sim$Ry, resulting in a basis size of about 3800. Local density approximation has been used for treatment of exchange and correlation effects \cite{Perdew1992,Elgazzar09}.

In Fig.\,S\ref{FigS4}(B) we highlight in particular the U spin-up character of the obtained surface projected bandstructure. Spin down character is significantly less pronounced within the shown energy range. There are several bands inside gaps of the bulk bandstructure, but only those around the $X$ point of the simple tetragonal Brillouin zone X$_{st}$ (see Fig.\,S\ref{FigS4}(C)), are sufficiently shallow to provide large effective masses.

We find a surface state (upper inset of Fig.\,S\ref{FigS4}(B)) which consists of two hybridized hole bands, forming an M-shaped feature close to the Fermi level. The dispersion relation found in our experiment (magenta line in the upper inset of Fig.\,S\ref{FigS4}(B) is compatible with the central part of the M-shaped feature. 

\clearpage

\section*{Section S5. Energy and position dependence of the features at the steps between terraces.}
\label{sec5}
\addcontentsline{toc}{section}{\normalfont Section S5. Energy and position dependence of the features at the steps between terraces.}

\begin{figure*}
	\begin{center}
	\centering
	\includegraphics[width=\textwidth]{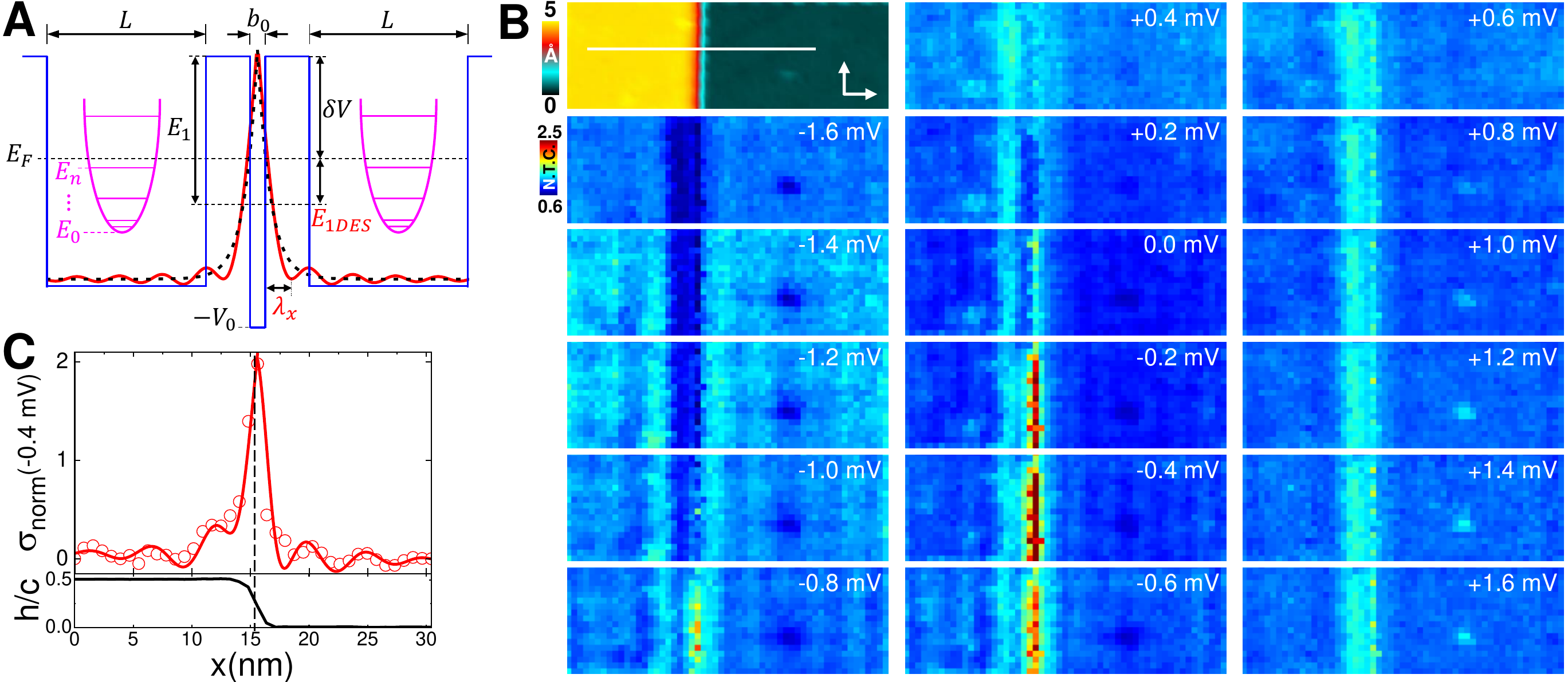}
	\end{center}
	\vskip -0.5 cm
	\caption{{\bf One dimensional edge states and tunneling conductance at the steps between terraces.} ({\bf A}) Schematic representation of the parameters used to describe the 1DES between two consecutive terraces of length $L$. We represent the quantized levels of the confined 2DHF on each terrace with pink color. Dashed black (continuous red) line represents the exponential behavior of the 1DES without (with) coupling with the quantized levels. ({\bf B}) Topography image along a step between two consecutive terraces is shown at the top left panel. The color scale is shown as a bar on the left, in \AA. White arrows provide the in-plane crystalline axis. The other panels are tunneling conductance images (color scale provided at the left) in the same field of view for different values of the bias voltage. ({\bf C}) Circles represent the tunneling conductance as a function of distance (referred to the tunneling conductance far from the edge at $x=50$ nm) at the bias voltage where the edge state appears $V\approx-0.4$ mV, $\sigma_{norm}(-0.4$ $mV)=\sigma(V=-0.4$ mV$,x)-\sigma(V,x=50$ $nm)$ along the white line in the top left panel of ({\bf B}). Continuous red line is the fit of the 1DES (Eq.(\ref{Eq5})) plus the quantized level model (Eq.(\ref{Eq1})). The bottom panel shows (black line) the STM height profile along the white line in the top left panel of ({\bf B}) in units of the c-axis lattice constant.}
	\label{FigS5}
	\end{figure*}

In order to analyze the one dimensional edge state (1DES) at the step between two terraces we use a 1D Dirac function like potential at the step, $V(x)=U_{0}\delta (x-x_{1DES})$, where $x_{1DES}$ is the position of the 1DES. We take $U_{0}=b_{0}V_{0}$, with $b_{0}$ the width of the potential well and $V_{0}$ the energy depth ($V_{0}<0$). We add a complex potential, $V(x)\rightarrow \left(U_{0}-iU_{1}\right) \delta (x-x_{1DES})$ to simulate the coupling of the edge state to the bulk of the crystal. A schematic representation of this model is shown in Fig.\,S\ref{FigS5}(A). Solving the Schr\"{o}dinger equation for $E<0$, we obtain 

\begin{equation}
g(E)=A\frac{e^{-|x-x_{1DES}|/\lambda_x}}{E-E_{1DES}+i\eta_{1DES}},
\label{Eq2}
\end{equation}

where $\lambda_x =\frac{\hbar ^{2}}{m^*}|U_{0}|^{-1}$ is the decay length with $m^*$ the effective mass; $E_{1DES}$ and $\eta_{1DES}$ are the energy position and the energy broadening of the 1DES given by

\begin{equation}
E_{1DES} =\delta V+E_1 =\delta V -\frac{m^*}{\hbar ^{2}}\left(U_{0}^{2}-U_{1}^{2}\right)
\label{Eq3}
\end{equation}
\begin{equation}
\eta_{1DES} =-\frac{m^*}{\hbar ^{2}}_{0}U_{1}
\label{Eq4}
\end{equation}

where $\delta V$ is the height of the well's potential barrier relative to the Fermi level. 

We can now fit the tunneling conductance at the 1DES using

\begin{equation}
g_{1DES}=A_{0} \frac{\eta_E e^{-\frac{\left|x-x_{1DES}\right|}{\lambda_x}}}{\left(E-E_{1DES}\right)^2+\eta_E^2}
\label{Eq5}
\end{equation}

Table\,S\ref{Tab1} shows the extracted fitting parameters $E_{1DES}$, $\eta_{1DES}$, $\lambda_x$ and $x_{1DES}$ for the four different terraces $L_1$ to $L_4$ from Fig.\,1(F,G). 

\begin{table}[ht]
\centering
\begin{tabular}{c c c c c c} 
\hline 
$L_n$ & $L [nm]$ & $E_{1DES} [mV]$ & $\eta_{1DES} [mV]$ & $\lambda_x [nm]$ & $x_{1DES}/a_0$ \rule{0pt}{2.6ex} \\ 
\hline \hline
$L_1$ & 2.00 & -0.65 $\pm$ 0.05 & 0.50 $\pm$ 0.10 & 0.9 $\pm$ 0.08 & 4.3 $\pm$ 0.4 \rule{0pt}{2.6ex}\\ 
$L_2$ & 5.50 & -0.48 $\pm$ 0.04 & 0.43 $\pm$ 0.05 & 0.9 $\pm$ 0.10 & 3.6 $\pm$ 0.5 \\ 
$L_3$ & 20.0 & -0.52 $\pm$ 0.02 & 0.52 $\pm$ 0.03 & 0.8 $\pm$ 0.05 & 3.8 $\pm$ 0.2 \\ 
$L_4$ & 38.5 & -0.45 $\pm$ 0.05 & 0.35 $\pm$ 0.05 & 0.8 $\pm$ 0.03 & 3.8 $\pm$ 0.2 \\ [0.5ex]
\hline 
\end{tabular}
\caption{{\bf 1DES parameters for terraces $L_1$ to $L_4$.} Values of the parameters extracted from the Eq.(\ref{Eq5}) used to fit the 1DES on the four different terraces $L_1$ to $L_4$ of Fig.\,1(G).} 
\label{Tab1} 
\end{table}

From Table\,S\ref{Tab1}, we see that the energy position and the energy broadening of the 1DES are independent of the terrace size with average values of $E_{1DES} = -0.525 \pm 0.04$ mV and $\eta_{1DES}=0.45\pm 0.06$ mV. We also see that all the spatial features are always at the same position with respect to the step, $x_{1DES}\approx 4.0 a_0$, being $a_0$  the in-plane constant lattice parameter, with a decay length $\lambda_x\approx 0.9$ nm $\approx 2 a_0$. The latter indicates that 1DES and 2DHF couple when the decay length reaches a few interatomic distances. With the extracted average values from Table\,S\ref{Tab1} for $\lambda_x$, $\eta_{1DES}$ and $E_{1DES}$, we obtain $U_{0}=5.4$ meV$\mathring{A}$, $U_{1}=0.38$ meV$\mathring{A}$ and $\delta V=3.1$ meV. 

We can analyze the 1DES through the tunneling conductance at a step (Fig.\,S\ref{FigS5}(B)). At low bias voltages we find a dip in the tunneling conductance a few nm at the upper side of the step. The dip fills with the 1DES at about $E_{1DES}$ and empties again at higher bias voltages. This shows that charge depletion close to the step opens a gap in the bandstructure. The gap is filled at the resonant energy of the 1DES. By normalizing the tunneling conductance to its shape far from the step (Fig.\,S\ref{FigS5}(C)), we can follow the decay of the 1DES into the quantum well states of the 2DHF with the model described above (Fig.\,S\ref{FigS5}(A)). The decay length is of order of the inverse of the wave vector of the 2DHF.

\clearpage

\section*{Section S6. Origin of $\varepsilon_{+}$ and results at point defects.}
\label{sec6}
\addcontentsline{toc}{section}{\normalfont Section S6. Origin of $\varepsilon_{+}$ and results at point defects.}

\begin{figure}
	\begin{center}
	\centering
	\includegraphics[width=\columnwidth]{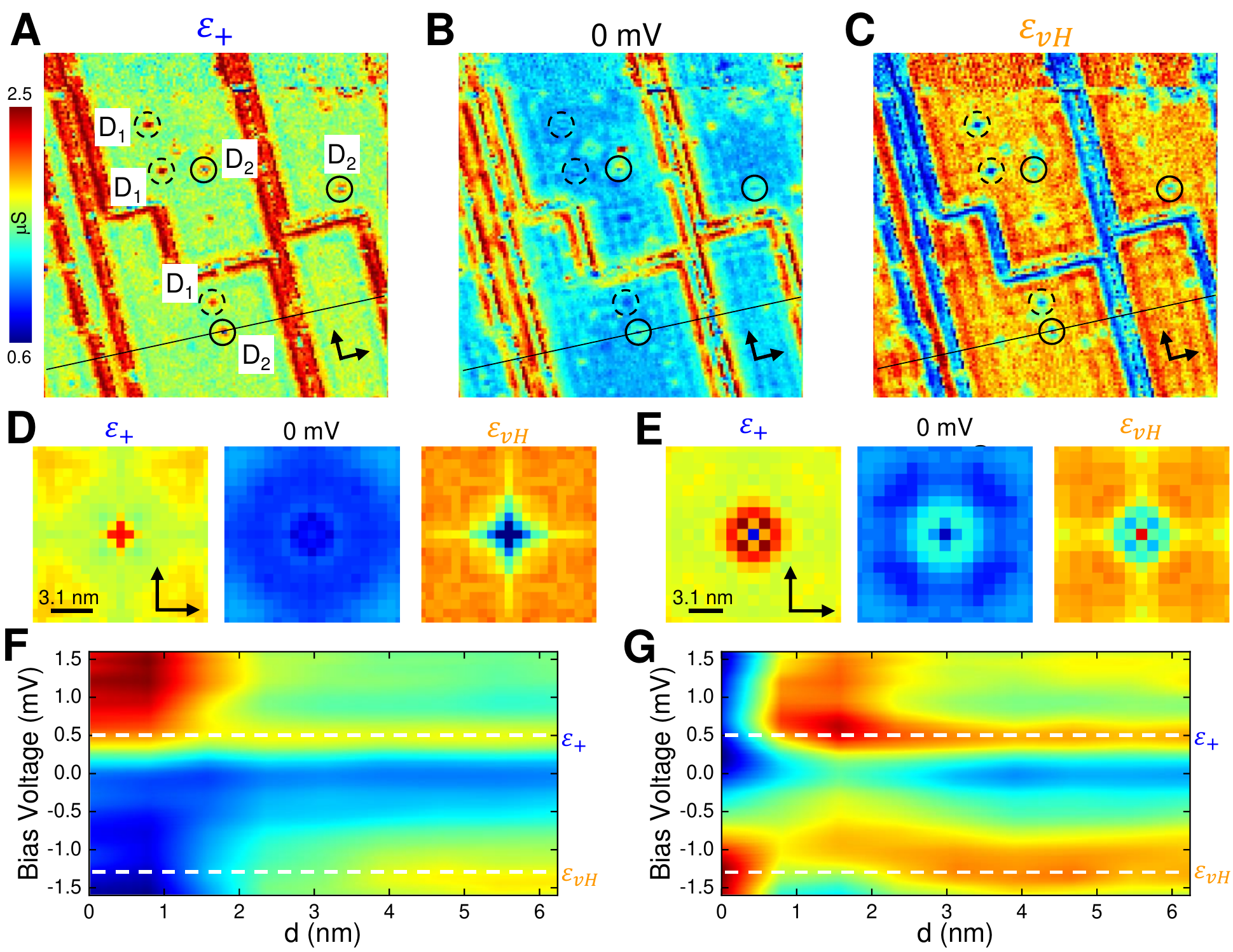}
	\end{center}
	\vskip -0.5 cm
	\caption{{\bf Tunneling conductance at atomic size defects.} ({\bf A-C}) Tunneling conductance maps at $\varepsilon_{+}$, zero bias and at $\varepsilon_{vH}$, respectively, in the same field of view as Fig.\,1(C). Two types of defects $D_1$ and $D_2$ are marked by dashed and filled circles. The black line on ({\bf A-C}) is the same line marked in Fig.\,1({\bf A}). ({\bf D-E}) Tunneling conductance maps at $\varepsilon_{+}$, zero bias and at $\varepsilon_{vH}$ in a field of view of $12.5\times12.5$ nm centered on the defects $D_1$ and $D_2$ respectively. ({\bf F-G}) Tunneling conductance as a function of the bias voltage and the distance for $D_1$ and $D_2$ respectively. White dashed lines provide $\varepsilon_{vH}$ and $\varepsilon_{+}$. Black arrows in ({\bf A-E}) are the [10] and [01] crystalline axis directions.}
	\label{FigS6}
	\end{figure}

We analyze here in more detail the tunneling conductance at defects. To this end, we plot the tunneling conductance at several terraces in Fig.\,S\ref{FigS6}(A-C) and focus on the behavior at defects. As discussed above, we can identify two kinds of defects, which we label here as D$_1$ and D$_2$. Defects D$_1$ are protusions in the images with height increases of around 40 pm, probably due to interstitial atoms located beneath the surface. D$_2$ are troughs of around 30 pm size, probably due to vacancies beneath the surface. The defects visibly affect the pattern formed by the 2DHF. We plot the tunneling conductance at $\varepsilon_+$, $0 mV$ and $\varepsilon_{vH}$ for both type of defects in Figs.\,S\ref{FigS6}(D-E). In Figs.\,S\ref{FigS6}(F-G) we show the spatial dependence of the the tunneling conductance along a crystalline axis for both type of defects. D$_1$ defects affect particularly the van Hove anomaly at $\varepsilon_{vH}$, with a strong suppression at the defect (Fig.\,S\ref{FigS6}(F)) which is more pronounced along a crystalline axis (square shape in Fig.\,S\ref{FigS6}(D) at $\varepsilon_{vH}$). The van Hove anomaly remains, by contrasts, at the D$_2$ defects. However, the feature at $\varepsilon_{+}$ is suppressed at the impurity site (Fig.\,S\ref{FigS6}(G)). This suggests that interstitial atoms (D$_1$) suppress the van Hove anomaly at $\varepsilon_{vH}$ and that vacancies (D$_2$) the feature at $\varepsilon_{+}$.

The local modification of the van Hove anomaly and of features close to $\varepsilon_{+}$ are similar to those observed in Th-doped URu$_2$Si$_2$ in a similar energy range and were discussed as Kondo holes\cite{Hamidian11,Morr2016}. Kondo holes are defects in the Kondo hybridization which locally modify the bandstructure. This suggests that both $\varepsilon_{vH}$ and $\varepsilon_{+}$ are related to the hybridization pattern close to the Fermi level.

Notice that the black line on Fig.\,S\ref{FigS6}(A-C) is the same line marked in Fig.\,1(A). Therefore, the reduction on the tunneling conductance signal close to $\varepsilon_{vH}$ and at around 40 nm distance in Fig.\,1(G) can be explained because of a defect of $D_1$ type.

\clearpage

\section*{Section S7. Model for the interplay between superconductivity and the 2DHF.}
\label{sec7}
\addcontentsline{toc}{section}{\normalfont Section S7. Model for the interplay between superconductivity and the 2DHF.}

Having realized the relevance of the features at $\varepsilon_{vH}$ and at $\varepsilon_{+}$ on the spatial dependence of the density of states, it is important to consider their influence on the superconducting properties. We notice that the quantum well states should influence superconducting properties in a similar way, but the occurrence of large features of $\varepsilon_{vH}$ and at $\varepsilon_{+}$ likely dominates the interaction with superconductivity. Furthermore, we consider several parallel conduction channels between the tip and the surface. For simplicity, we take into account tunneling into the 2DHF and into the feature of largest size at $\varepsilon_{vH}$. The first channel $t_1$ connects the tip with the 2DHF. The 2DHF is superconducting by proximity from the bulk superconductor, which we model using a coupling $t_s$. With the second channel, $t_2$, we connect the tip to bulk states. We furthermore consider a hybridization between the 2DHF  and the bulk states at the van Hove feature $\varepsilon _{vH}$, $t_{vH}$ and the feature at $\varepsilon _{+}$, $t_+$.

We then write the retarded Green function $\widehat{G}^r$,

\begin{equation}
\begin{aligned}
\widehat{G}^r=\left( 
\begin{array}{cc}
\widehat{M}_{2D}&\widehat{t}_{vH}  \\ 
\widehat{t}_{vH} &\widehat{M}_{vH} 
\end{array}\right) ^{-1}
\end{aligned}
\label{Eq6}
\end{equation}

with
\begin{equation}
\begin{aligned}
\begin{split}
\widehat{M}_{2D}&=\left( 
\begin{array}{cc}
E-E_{2DHF}-\frac{t_+^2}{E-E_+}+\frac{E+i\eta}{\Omega}\bar{t}_s&\frac{\Delta}{\Omega}\bar{t}_s \\ 
\frac{\Delta}{\Omega}\bar{t}_s &E+E_{2DHF}^*-\frac{t_+^2}{E+E_+^*}+\frac{E+i\eta}{\Omega}\bar{t}_s
\end{array}
\right)\\
\widehat{M}_{vH}&=\left( 
\begin{array}{cc}
E-E_{vH} & 0 \\ 
0            & E+E^*_{vH}
\end{array}
\right)\\
\widehat{t}_{vH}&=\left( 
\begin{array}{cc}
t_{vH} & 0 \\ 
0          & -t_{vH}
\end{array}
\right)\\
E_{j}&=\varepsilon _{j}- i\frac{\Gamma _{j}}{2}, (j=2DHF,vH,+)\\
\bar{t}_s&=\frac{t_s^2}{W}\\
\Omega&=\sqrt{\Delta ^{2}-(E+i\eta)^{2}}
\end{split}
\end{aligned}
\label{Eq7}
\end{equation}

where $\varepsilon _{2DHF}$ is the energy associated to the 2DHF and includes the shift of energy due to HO and Fano resonance, $W$ is an energy scale related to the normal density of states at the Fermi level by $\rho(E_F)=1/(\pi W)$, $\Delta$ is the superconducting gap and $\eta$ is a small energy relaxation rate. We
have added the self energies $i\Gamma _{j}/2,$ ($j=2DHF,$ $vH,$ $+$) , with $%
\Gamma _{j}\ $\ the width of the resonance $j$. 

The differential conductance is calculated as

\begin{equation}
\sigma (V)=\sigma _{0}\int T\left( E\right) \left( -\frac{df(E-eV,T)}{d\left(
eV\right) }\right) dE
\label{Eq8}
\end{equation}

with

\begin{eqnarray}
T\left( E\right)  &=&-\frac{1}{\pi }Im\left( G^r_{11}\left( E\right) t_{1} ^{2}+G^r_{33}\left( E\right) t_{2} ^{2}+\right.   \nonumber \\
&&\left. t_{1}t_{2}\left(G^r_{13}\left( E\right) +G^r_{31}\left( E\right)
\right) \right) 
\label{Eq9}
\end{eqnarray}

$f(E,T)$ the Fermi–Dirac distribution at the energy $E$ and temperature $T$ and $\sigma _{0}=\frac{2e^2}{h}$ the quantum of conductance (with $h$ being Planck's constant and $e$ the elementary charge). Notice that $T$ is the transmission, not the density of states used often to discuss STM measurements in superconductors. Notice also that we take into account tunneling into the 2DHF ($G^r_{11}$) and into $\varepsilon_{vH}$ ($G^r_{33}$), with mixed contributions ($G^r_{31}$ and $G^r_{13}$).

To fit the tunneling conductance curves between $0.25K \leq T \leq 1.8 K$ shown in Fig.\,3(B-C) we have subtracted an asymmetric background (Fig.\,1(E) of the main text). In Table S\ref{Tab2} we show the parameters used to obtain the tunneling conductance (shown as black lines in Fig.\,3(C) of the main text) from the Eq.\,(\ref{Eq8}). We fix use $\eta=0.018\, meV$, $\varepsilon_{2DHF}=12\, meV$, $\Gamma_{2DHF}=1\, meV$, $\Gamma_{vH}=0.55\, meV$ and $\Gamma_+=0.14\, meV$. We see that the superconducting lifetime itself is practically negligible $\eta=0.018$\, meV$\ll$ $\Delta=0.2$\, meV. The large zero bias conductance is not only due to the incomplete coupling to the bulk, as $t_s$ is close to one. The features at $\varepsilon_+$ and at $\varepsilon_{vH}$, $\Gamma_+$ and $\Gamma_{vH}$, provide Lorentzians that considerably smear the superconducting density of states and produce a finite tunneling conductance at zero bias.

\begin{table}[ht]
\centering
\begin{tabular}{c c c c c c c c c} 
\hline 
$T [K]$ & $t_1$ & $t_2$ & $t_+$ & $t_s$ & $t_{vH}$ & $\varepsilon_+ [meV]$ & $\varepsilon_{vH}  [meV]$ & $\Delta [meV]$ \rule{0pt}{2.6ex} \\ 
\hline \hline

0.25	&	0.89	&	0.24	&	0.70	&	0.82	&	0.80	&	0.70	&	-1.25 & 0.20\rule{0pt}{2.6ex}\\ 
0.35	&	0.97	&	0.24	&	0.70	&	0.89	&	0.87	&	0.67	&	-1.25 & 0.20\\ 
0.40	&	0.91	&	0.24	&	0.74	&	0.83	&	0.82	&	0.62	&	-1.25 & 0.20\\ 
0.45	&	0.90	&	0.22	&	0.61	&	0.83	&	0.81	&	0.66	&	-1.20 & 0.20\\ 
0.50	&	0.88	&	0.24	&	0.74	&	0.81	&	0.79	&	0.66	&	-1.25 & 0.20\\ 
0.55	&	0.91	&	0.22	&	0.73	&	0.84	&	0.82	&	0.63	&	-1.23 & 0.20\\ 
0.60	&	1.00	&	0.25	&	0.78	&	0.92	&	0.90	&	0.67	&	-1.25 & 0.20\\ 
0.70	&	1.05	&	0.27	&	0.90	&	0.97	&	0.95	&	0.68	&	-1.20 & 0.20\\ 
0.80	&	0.88	&	0.26	&	0.75	&	0.81	&	0.79	&	0.64	&	-1.30 & 0.18\\ 
0.90	&	0.93	&	0.25	&	0.73	&	0.85	&	0.83	&	0.64	&	-1.30 & 0.18\\ 
1.00	&	1.17	&	0.25	&	0.90	&	1.07	&	1.05	&	0.67	&	-1.30 & 0.18\\ 
1.10	&	0.93	&	0.26	&	0.75	&	0.85	&	0.83	&	0.64	&	-1.30 & 0.16\\ 
1.20	&	1.20	&	0.24	&	0.85	&	1.00	&	0.98	&	0.70	&	-1.20 & 0.16\\ 
1.30	&	1.05	&	0.24	&	0.87	&	0.97	&	0.95	&	0.70	&	-1.20 & 0.14\\ 
1.40	&	1.05	&	0.23	&	0.85	&	0.97	&	0.95	&	0.70	&	-1.20 & 0.12\\ 
1.70	&	1.00	&	0.23	&	0.87	&	0.92	&	0.90	&	0.70	&	-1.20 & 0.00\\ 
1.80	&	1.00	&	0.24	&	0.92	&	0.92	&	0.90	&	0.73	&	-1.20 & 0.00 \\ [0.5ex]
\hline 
\end{tabular}
\caption{{\bf Fitting parameters from the tunneling conductance in the range 0.25 K$\leq$T$\leq$1.8 K.} Values of the parameters extracted from the Eq.\,(\ref{Eq8}) used to fit the tunneling conductance curves shown as black lines in Fig.\,3({\bf B}-{\bf C}) of the main text.} 
\label{Tab2} 
\end{table}

\clearpage

\bibliographystyle{Science}